\documentclass[prl,aps,twocolumn,showpacs,nofootinbib,nobibnotes,superscriptaddress]{revtex4}
\usepackage{epsfig}
\usepackage{amssymb}
\begin{document}
\title{
Neutrino Mass Limit from Galaxy Cluster Number Density Evolution}
\date{\today}
\author{Tina Kahniashvili}
\email{tinatin@phys.ksu.edu}
\affiliation{Department of Physics, Kansas State University,
116 Cardwell Hall, Manhattan, KS 66506, USA}
\affiliation{
Center for Plasma Astrophysics, Abastumani Astrophysical Observatory,
2A Kazbegi Ave, GE-0160 Tbilisi, Georgia}
\author{Eckhard von Toerne}
\email{evt@phys.ksu.edu}
\affiliation{Department of Physics, Kansas State University,
116 Cardwell Hall, Manhattan, KS 66506, USA}
\affiliation{
Physikalisches Institut, Universit\"{a}t Bonn,
Nussallee 12, D-53115 Bonn, Germany}
\author{Natalia A. Arhipova}
\email{arna@lukash.asc.rssi.ru}
\affiliation{Astro Space Center, P.~N.~Lebedev Physical Institute,
84/32 Profsoyuznaya, Moscow 117997, Russia}
\author{Bharat Ratra}
\email{ratra@phys.ksu.edu}
\affiliation{Department of Physics, Kansas State University,
116 Cardwell Hall, Manhattan, KS 66506, USA}

\date{March 2005 \hspace{0.3truecm} KSUPT-05/1}
\begin{abstract}
Measurements of the evolution with redshift of the number density of
massive galaxy clusters are
used to constrain the energy density of massive neutrinos
 and so the sum of neutrino masses $\sum m_\nu$. We consider
a spatially-flat cosmological model with cosmological constant,
cold dark matter, baryonic matter, and massive neutrinos.
Accounting for the uncertainties in the measurements of the
 relevant cosmological parameters
we obtain a limit of
$\sum m_\nu$ $<$ $2.4$ eV (95 \% C.L.).
\end{abstract}

\maketitle

 Constraints on neutrino masses are of great interest for particle
physics as well as for cosmology, and thus attract a lot of
scientific attention (for recent reviews see
Refs.~\cite{dolgov,wong04,eidelman04,elgaroy04}). 
Current upper limits on
the sum of neutrino masses, $\sum m_\nu$,  from cosmological
structure formation  data \cite{elgaroy02,elgaroy03}, cosmic
microwave background (CMB) fluctuation data
\cite{pierpaoli03,ichi}, or combined CMB + large--scale structure
data \cite{barger1,crotty,lesgourges,seljak04,ales}, 
 are of order an eV \cite{fogli}
(for various limits see Table 1 of Ref.~\cite{elgaroy04}). The
number of neutrino species can  be constrained from Big Bang
nucleosynthesis  or by using CMB and large--scale structure data  
\cite{barger,crotty}.

High energy physics experiments also constrain neutrino masses, and
have measured the number of light neutrino species with high precision
\cite{eidelman04}.
Direct searches for neutrino mass effects in beta decays yield limits
in the region of several eV, but the sum over all neutrino masses is
almost unconstrained by beta decay and other experiments, mainly due to the
weak limit on the tau-neutrino mass.
The measurement of neutrino oscillations on the other hand constrains
the differences between the squared masses of the neutrino mass eigenstates
$\Delta m^2$. With the justified assumption that neutrino masses are
non-negative and for mass splittings
$\Delta m_{\odot }^2 \approx 7\times 10^{-5}$ eV$^2$ and $\Delta
 m^2_{\rm atm} > 1.3\times 10^{-3}$ eV$^2$ \cite{eidelman04}, we obtain
$\sum m_\nu$ $>0.04$ eV if the solar mass splitting is between the
highest and second highest mass eigenstates ($\sum m_\nu$ $>0.07$ eV if the
atmospheric  mass splitting is between the two highest states). Results from
the LSND collaboration yield  a larger lower limit on $\sum m_\nu$,
and must be considered if
 confirmed by the MiniBooNE experiment \cite{Ray:2004hp},
which is currently taking data.

In this paper we use the dependence of galaxy cluster number density
evolution on the massive neutrino energy density parameter $\Omega_\nu$
to set a limit on $\sum m_\nu$. We consider
the standard spatially-flat 
 $\Lambda$CDM Friedmann--Lema\^\i tre-Robertson-Walker 
 cosmological spacetime model
with baryons, cold dark matter (CDM), massive neutrinos, and
a non-zero cosmological constant $\Lambda$ (for a recent
 review see Ref.~\cite{peebles03}).
To compute the cluster number density as a function of
redshift $z$ we use the Press--Schechter approach
\cite{PrSch74,bahcall98} as modified by
Sheth and Tormen (ST) \cite{sheth}.\footnote{In the
  following we use PS to refer to the un-modified Press--Schechter
approach.}
This approach makes use of the mass function  $N(M>M_0)$
 of clusters (cluster number density as a function of
cluster mass $M$ greater than a fiducial  mass $M_0$),
which depends
on cosmological model parameters \cite{nfw,cole97,white,white1}.
In particular, it is  very
sensitive to the matter density parameter
 $\Omega_M$ ($=\Omega_b+\Omega_{\rm{cdm}}+\Omega_\nu$, 
 where $\Omega_b$ and $\Omega_{\rm{cdm}}$ are the density parameters of
 baryons and CDM, respectively) and the value of
$\sigma_8$ (the r.m.s. amplitude of density fluctuations smoothed over
a sphere of $8 h^{-1}\mbox{Mpc}$ radius, where $h$
is Hubble constant in units of 100 km~s$^{-1}$Mpc$^{-1}$)
\cite{BodeBac03}.
The observationally-viable ranges of these
two parameters are related (for recent
reviews see Refs.~\cite{tegmark04,seljak04,cole05}),
 and a current version of
 the relation between
$\Omega_M$ and $\sigma_8$ is given in  Table 5 of
Ref.~\cite{tegmark04}. The parameter $\sigma_8$ is determined by the
 matter fluctuation
power spectrum which is sensitive  to
$\Omega_\nu$ \cite{elgaroy03,wong04}. This is because 
the neutrinos are light particles and have a much larger free streaming 
path length\footnote{For ultrarelativistic particles the 
 free streaming path length  
is equal to Hubble radius. After they became non-relativistic, 
 particle velocities are redshifted away adiabatically and so the 
free streaming path length grows only slowly or decreases   
\cite{crotty}.} than the
CDM particles. Gravitational
instability is therefore unable to confine the neutrinos on small and
intermediate length scales, resulting in a suppression of
 small- and intermediate-scale power. See
  Fig.~6 of
Ref.~\cite{elgaroy04} for
$\sigma_8$ as a function of $\Omega_\nu$ for models
normalized to the
Wilkinson Microwave Anisotropy Probe (WMAP) data. 

 Neutrinos are weakly interacting and this characterises how they affect 
cosmology. When  
$m_\nu > 10^{-3}$ eV neutrinos are non-relativistic today \cite{crotty}  
and thus behave like  a hot component of dark matter. 
The presence of even a small fraction of massive neutrinos (hot dark 
matter)  $f_\nu 
\equiv \Omega_\nu/\Omega_M$, of order 10-20 \%, requires a
smaller  value for the cosmological constant in comparison to a
pure $\Lambda$CDM model,  while other cosmological parameters are
largely unaffected \cite{elgaroy03}.
 This is because a smaller
$\Omega_\Lambda$ results in a larger $\Omega_M$ and
hence a faster fluctuation
growth rate, which compensates for the reduction of  small- and
 intermediate-scale power caused by the neutrinos.
Neutrino free-streaming suppression of the linear\footnote{The effect
of massive neutrino infall into CDM halos is studied
in Ref.~\cite{abaz}. They found that three degenerate-mass neutrinos with
$\sum m_\nu \sim 2.7$ eV alters the non-linear matter power spectrum
by about  1 \%.} growth of density
perturbations on small and intermediate scales results in only a
fraction of matter of order $1-f_\nu$ being involved in
gravitational clustering \cite{bond}.  
 Being an integral over the 
power spectrum,  $\sigma_8$ depends sensitively on $\Omega_\nu$. 
This makes the cluster number density evolution with redshift  
 very sensitive 
to the value of $\Omega_\nu$ \cite{val,Arkh2001}, and so to the value of 
$\sum m_\nu$ (since $\sum m_\nu = 94 \Omega_\nu h^2 $ eV \cite{eidelman04}). 

 As with other cosmological tests, the cluster number density evolution 
test for neutrino masses requires fixing  the range of some 
 cosmological parameters. This may be viewed as a 
 choice of priors; see Ref.~\cite{elgaroy03} 
for a detailed duscussion of priors in the context of deriving neutrino 
 mass limits from cosmological data. 
 The parameter ranges we consider in this computation are picked
as follows. 
 Based on Hubble Space Telescope (HST) measurements of
the Hubble constant \cite{HST}, we use 
 $h = 0.71 \pm 0.07$ (one
standard deviation limit).\footnote{ An analysis of all available
measurements of the Hubble constant results in the more
restrictive, but HST consistent, estimate $h=0.68 \pm 0.07$ (two
standard deviation range) \cite{gang-h}.} We assume adiabatic
density perturbations with primordial power spectral index close
to scale-invariant,  
 $n = 0.98 \pm 0.02$ (1-$\sigma$ range)
\cite{seljak04,tegmark04}. Concerning the value of the matter
density parameter $\Omega_M$, there is  evidence for $\Omega_M
\in (0.2, 0.35)$  from different data such as  Type Ia supernovae
\cite{SNIa}, WMAP  \cite{spregel}, Sloan Digital Sky Survey (SDSS)
and 2 degree Field Galaxy Redshift Survey (2dFGRS) galaxy
clustering \cite{tegmark04,cole05}, and galaxy cluster gas mass
fraction evolution \cite{32b}. For a summary see Ref.~\cite{32a}, 
who find 
 $0.2 \lesssim \Omega_M \lesssim 0.35$ at two standard
deviations; in our computation we use this as a one standard
deviation range.
 We choose
$\sigma_8 \in (0.77, 1.11)$ as the two standard deviation range;
 for a discussion
see Sec. 3.1 of Ref.~\cite{Viel:2004np}.  For the baryon density
parameter we use from Big Bang nucleosynthesis
$\Omega_b h^2 \in (0.018, 0.022)$.\footnote{This is more 
 consistent with estimates from WMAP data and from Big Bang nucleosynthesis
  using the mean of the primordial
deuterium abundance measurements, but it is significantly larger than an
 estimate based on helium and lithium abundance measurements, 
 see, e.g., Ref.~\cite{peebles03}. } 
We work in  a spatially-flat model which requires
$1-\Omega_\Lambda=\Omega_b +\Omega_{\rm {cdm}}
+\Omega_\nu=\Omega_M$. The prior on $\Omega_b$  
is less important than  the priors on $\sigma_8$, 
 $\Omega_M$, $h$ or $n$ \cite{elgaroy03}.

   The paucity of galaxy cluster evolution data makes it inappropriate to
use
data analysis techniques based on $\chi^2$ fits. Reviewing such data-poor
situations in the literature
one finds either modified $\chi^2$ fits
assuming gaussian errors on the logarithm of the observed cluster number
density  or likelihood approaches with Poisson
errors based on the number of observed clusters.
In our analysis we use a likelihood approach
and we define our likelihood function by 
$\mathcal{L}=\prod_i e^{-\mu_i} \mu_i^{k_i} /k_i!$.  Here $k_i$ is the
number of
observed clusters with mass greater that $M_0$,
 in the $i^{\rm {th}}$ redshift  bin (centered at redshift $z_i$ 
 and of width $\Delta z$)  and
$\mu_i (M>M_0, z_i, \Delta z ) = \int_{z_i}^{z_i +\Delta z} \:
N_{\rm{pred}}(M\!>\!M_0, z)\:
{\rm d}z / (\Delta z \: \alpha_i) $ is the predicted
cluster number in this  bin. $N_{\rm{pred}}(M>M_0, z)$ is the predicted
cluster number density.  The normalization factor $\alpha_i$, 
 which has dimensions of inverse volume, is
the detection efficiency defined by $N_{\rm obs}(M>M_0, z_i) = 
\alpha k_i $ where
$N_{\rm obs}(M>M_0, z_i)$ is the observed cluster number density in $i^{\rm{th}}$ redshift bin.

   The predicted cluster number density $N_{\rm{pred}}(M>M_0, z)$
depends on the cosmological model considered. An important
characteristic of a cosmological
model is the  linear energy density perturbation power spectrum $P
(k, z)$. This  is sensitive to the values of the
 cosmological parameters $h$, $\Omega_M$, and $\Omega_b$, 
 as well as to
 the density parameter  of each  dark matter component,
i.e., $\Omega_{\rm {cdm}}$ and $\Omega_\nu$ (although the requirement of flat
 spatial hyperspaces can be used to 
eliminate the dependence on $\Omega_{\rm{cdm}}$), and to
additional parameters discussed below. This wavenumber-space
two-point correlation function of density perturbations, $P (k,
z)$, at given redshift $z$, is determined by the power spectrum of
initial perturbations $P_0(k)$, the energy density perturbation
transfer function $T(k,z)$, and the perturbation growth rate
$D(z)$. The functions $T(k,z)$ and $D(z)$ depend on model
parameters and describe the evolution of density 
inhomogeneities \cite{textbook}. To compute $P (k, z)$ we assume
a close to scale-invariant (Harrison-Peebles-Yu-Zeldovich)
post-inflation energy density perturbation power spectrum $P_0(k)
\propto k^n$, with $n \sim 1$, and we use a semi-analytical
approximation
 for the transfer function $T(k,z)$ in the
$ \Lambda$CDM model with three species of equal-mass neutrinos
 \cite{EisHu97}.\footnote{The effects of neutrino
 mass differences are  irrelevant for
our considerations,
since  if $\sum m_\nu >0.4$ eV the mass eigenvalues are essentially
degenerate \cite{elgaroy04}, while if $\sum m_\nu <0.4$ eV the mass
differences do not much affect  cluster number density
evolution \cite{crotty}.}
For the growth rate $D(z)$
we use Eqs.~(2)--(4) of Ref.~\cite{Arkh2001},
 which are based on results from Refs.~\cite{KofStar85}.
In summary, in our model the free parameters are $n$, $h$,
$\Omega_M$, $\sum m_\nu$ (or $\Omega_\nu$), and
$\sigma_8$ (which fixes the 
normalization of the power spectrum, as discussed below).

The cluster mass function at redshift $z$ is 
$N(M>M_0, z)=\int_{M_0}^\infty dM ~n(M, z)$, where $n(M, z)dM$ is the
comoving number density of collapsed objects with mass lying in
the interval $(M, M+dM)$. In the PS approach the cluster mass
function  is
 determined by $\sigma(R,z)$ 
 the r.m.s.
 amplitude of density
fluctuations  smoothed over a sphere of  radius
$R=(3M/4\pi\rho_M)^{1/3}$, where $\rho_M$ is the mean matter
density \cite{PrSch74}. The function $n(M, z) $ is a universal
function of the peak height $\delta_C/\sigma(R)$, where
$\delta_C=1.686$.  For  gaussian  fluctuations,
\begin{equation}
n(M,z) \propto
 \frac{\delta_C}{\sigma^{2}(R,z)} \left| \frac{d\sigma(R,z)}{dM} \right|
 \mbox{exp} \left[-\frac{\delta^{2}_{C}}{2\sigma^{2}(R,z)}\right]\;,
\label{N1}
\end{equation}
see, e.g.,  Eq.~(1) of Ref.~\cite{bahcall98}. The evolution of
the cluster mass function  is determined by the $z$ dependence of
$\sigma(R,z)$. Now  $\sigma^{2}(R,z)$ is related to the power
 spectrum $P(k,z)$ through \begin{equation}
\sigma^{2}(R,z)= \frac{1}{2 \pi^{2}}\int\limits_{0}^{\infty}
P(k,z)|W(kR)|^{2}k^{2}dk,
\label{sigma}
\end{equation}
where $W(kR)$ is the Fourier transform of the top-hat window
function, $ W(x)={3}(\sin{x}- x\:\cos{x})/x^3 $.
 Numerical computation results for $n(M, z)$ are not accurately fit by the 
PS expression of Eq.~(\ref{N1}), see Refs.~\cite{sheth,jen,numeric}.
 Several more accurate modifications of $n (M, z)$
have been proposed, see Refs.~\cite{sheth,modification,jen}. Here we use the ST
modification \cite{sheth}, as defined in Eq.~(5) of
Ref.~\cite{white1},
\begin{eqnarray}
n(M, z) \propto \frac{\delta_C}{\sigma^{2}(R,z)} \left| \frac{d\sigma(R,z)}{dM} \right|\left[ 1+
\left(\frac{a\delta^{2}_{C}}{\sigma^{2}(R,z)}\right)^{-p} \right]
\nonumber\\\times 
\left(\frac{a\delta^{2}_{C}}{\sigma^{2}(R,z)}\right)^{-1/2}
\mbox{exp}\left[-\frac{a\delta^{2}_{C}}{2\sigma^{2}(R,z)}\right]~,
\label{f}
\end{eqnarray}
where the parameters $a=0.303$ and $p=0.707$ are fixed by fitting to the
numerical results (for the PS case $a=1$ and $p=0$)
\cite{sheth,white1}. With this choice of parameter values the
mass of collapsed objects in Eq.~(\ref{f})  must be defined using
a fixed over-density contrast with respect to the background
 density  $\rho_M$ \cite{jen,white,white1,BodeBac03}, and this
requires accounting for the mass conversion between $M_{180b}$ and
$M_{200c}$ \cite{white,white1}.\footnote{The mass of a collapsed  
object is defined 
 with respect to the Einstein-de Sitter critical
 density $\rho_{\rm{cr}}=3H^2/(8\pi G)$, as the mass within a radius 
$R_{\Delta c}$,
inside of which the mean interior density is $\Delta$ times
 the critical density $\rho_{\rm{cr}}$. Assuming a $\rho(R)$
density profile, $\int_0^{R_{\Delta c}} \rho(R)R^2dR =
\Delta\rho_{cr} R_{\Delta c}^3/3$. $M_{200c}$, which corresponds to 
 an over-density $\Delta=200$ with respect to the critical density, 
 is a common 
definition of the virial mass of the cluster \cite{nfw,white1}, 
 while $M_{180b}$ corresponds to the mass
within a sphere of radius $R_{180b}$ inside which the mean density is 
$180$ times the background density $\rho_M$  
($R_{180b}=R_{54c}$ for $\Omega_M=0.3$) \cite{white1}. For details of 
 mass conversion
assuming a Navarro-Frenk-White  density profile  \cite{nfw} 
see \cite{white1}.} Such a conversion depends on
 cosmological parameters, see  Fig.~1 of Ref.~\cite{white}; we use  
an analytical extrapolation  of this  figure to do 
 the conversion for $\Omega_M \in (0.2, 0.35)$.

 Our analysis is based on data from a compilation of massive clusters (with
$M > M_{0} = 8 \times 10^{14} h^{-1} M_{\odot}$, where $M_{0}$
here corresponds to the mass within a comoving radius of $1.5
h^{-1}\mbox{Mpc}$ and $M_\odot$ is the solar mass) observed at
redshifts up to $z \approx 0.8$, \cite{bahcall98,BodeBac03}, 
\begin{eqnarray}  
N(M > 8 \times 10^{14} h^{-1} M_{\odot})   h^{-3}\mbox{Mpc}^{3}~~~~~~~~~~~~~~~~~~~~~~~~~
& \cr   
  \nonumber\\
~~~~~~~~~~= \cases{ 
 1.1{+1.1\atop-0.7}\times 10^{-7},  & \mbox{at}~~~$z=0.0-0.1$, \cr
 1.7{+1.7\atop-1.1} \times 10^{-8}, & \mbox{at}~~~$z=0.3-0.5$, \cr 
 1.4{+1.1\atop-0.9} \times 10^{-8}, & \mbox{at}~~~$z=0.5-0.65$, \cr
 1.4{+1.4\atop-1.1} \times 10^{-8}, & \mbox{at}~~~$z=0.65-0.9$. \cr} 
\label{data}
\end{eqnarray}
Here the  errors in $N$ are from counting uncertainty  and are 
 automatically included in our Poisson likelihood analysis. 
 The bin width in $z$ is chosen to reflect the uncertainty in the 
 redshift measurement.
 \begin{figure}[htbp]
\centerline{\includegraphics[scale=0.42,clip=true]{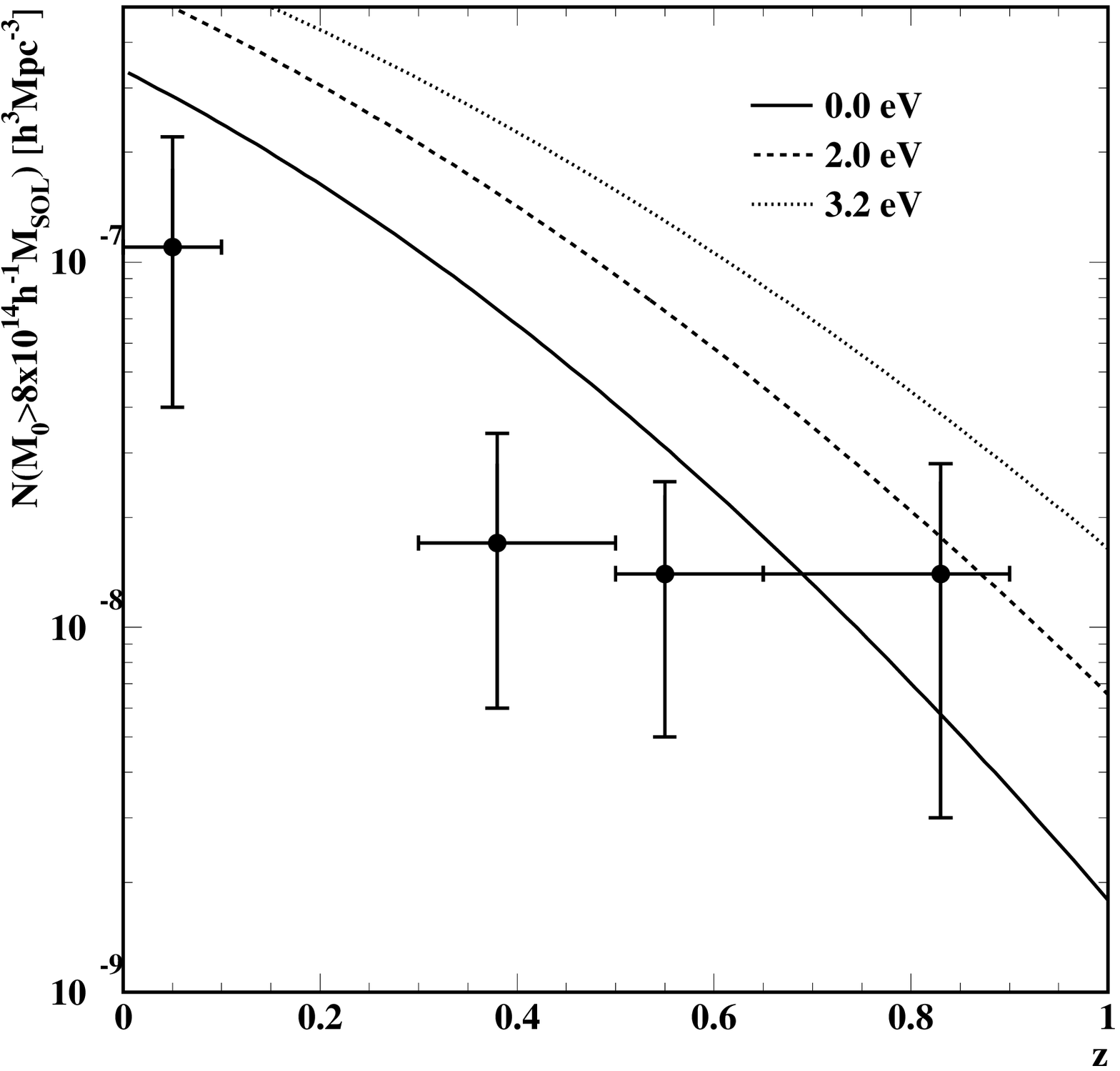}}
\caption{The curves show the number density evolution of massive clusters
  ($M > 8 \times 10^{14} h^{-1} M_{\odot}$)
for models in which the sum of neutrino masses $\sum m_\nu$=0,~2,~3.2 eV 
(from bottom to top). The other five parameters $\Omega_M$, $\Omega_b$,
  $\sigma_8$, $n$, and $h$ are set at the center of the scan
  interval. The crosses show 
  the observational data of Eq.~(\ref{data}) 
 in four redshifts bins with 1-$\sigma$ Poisson error bars.}
\label{fig1}
\end{figure}

\begin{figure}[htbp]
\centerline{\includegraphics[scale=0.42,clip=true]{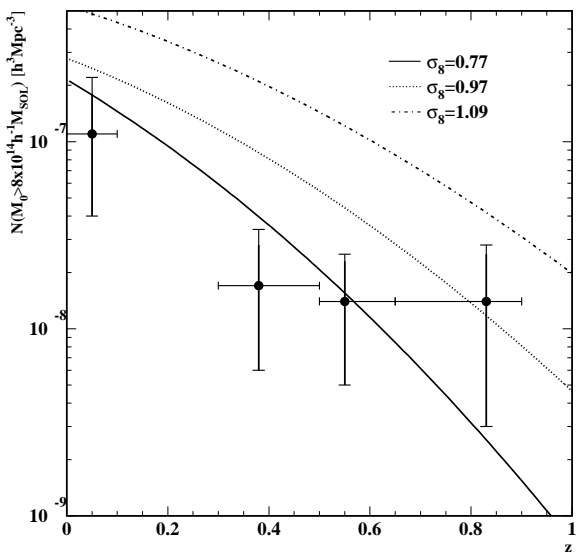}}
\caption{Variation of cluster number density evolution 
$N(M>8 \times 10^{14} h^{-1} M_{\odot}, z)$ as a function of 
$\sigma_8$ with parameters kept at the same values 
as in Figure \ref{fig1} except for $\sigma_8$ which is varied and 
$\sum m_\nu$ which is kept at 
zero. The observational data are the same as those shown in Figure 1.}
\label{fig2}
\end{figure}

We compute the likelihood from the data using Poisson errors
and the predicted number of clusters in each bin, and perform a maximum
likelihood fit over a discretized parameter space.
We compare the observed cluster number density  evolution of 
 massive clusters with
$M > 8 \times 10^{14} h^{-1} M_{\odot}$ to model predictions for different 
 values of $\sum m_\nu$ (see Figure \ref{fig1} for some examples) 
and for each 
 value of $\sum m_\nu$ we marginalize the likelihood by integrating
 over the parameter space 
$(\Omega_M, \Omega_b, h, n, \sigma_8)$ with Gaussian weighting. 
Figure \ref{fig2} shows the dependence of cluster number density evolution 
 on $\sigma_8$. Here  
 $\sum m_\nu=0$ and other four parameters $(\Omega_M, 
\Omega_b, h, n)$ are set at the center of the scan interval. 

\begin{figure}[htbp]
\centerline{\includegraphics[scale=0.42,clip=true]{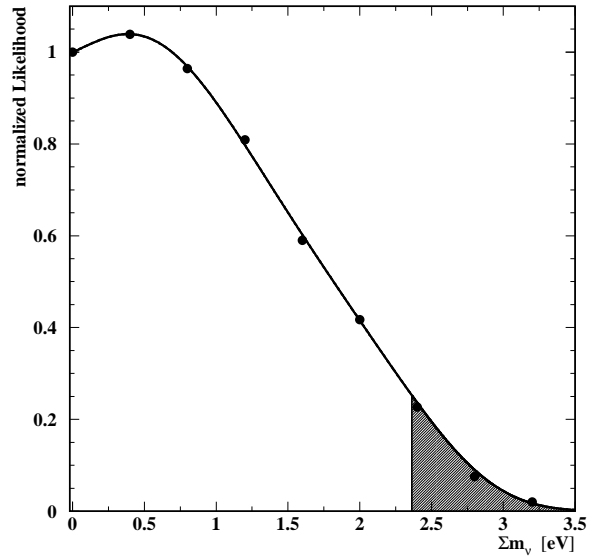}}
\caption{The marginalized (integrated) likelihood $\mathcal{L}/\mathcal{L}_{max}$ as a
  function of $\sum m_\nu$, accounting for the observational
  uncertainties in $\Omega_M$, $\Omega_b$, $\sigma_8$, $n$, and $h$. 
The shaded region is excluded at 95\% confidence level.
The likelihood function has been scaled for display purposes
to a value of one at $\sum m_\nu=0$.}
\label{fig3}
\end{figure}

\begin{figure}[htbp]
\centerline{\includegraphics[scale=0.42,clip=true]{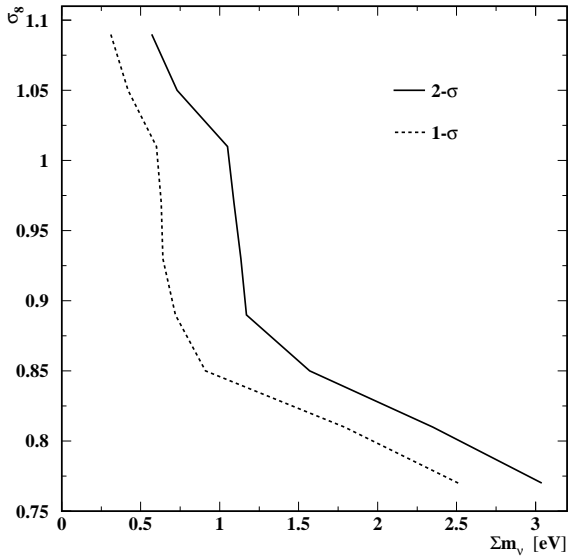}}
\caption{The curves show the two-dimensional likelihood in 
$(\sigma_8,\sum m_\nu)$. Continuous line is the  2-$\sigma$ contour and the 
 dashed line is the 
1-$\sigma$ contour. 
Lower values of $\sigma_8$ favor a higher neutrino
mass, for example, at $\sigma_8=0.77$ the most likely neutrino mass
value is $\sum m_\nu=1.3$ eV (also see Ref.~\cite{al} ) 
 whereas at high $\sigma_8$ values only upper limits 
 on $\sum m_\nu$ apply.}
\label{sig8}
\end{figure}

The likelihood as a function of $\sum m_\nu$ is shown
in Figure \ref{fig3}. Low values of $\sum m_\nu$ are favored but we find no
preference for a non-zero value. 
 We obtain an upper limit of 
$\sum m_\nu < 2.37$ eV (95 \% C.L.), accounting for the uncertainties in 
all five cosmological parameters $\Omega_M$, $n$, $h$, $\sigma_8$, and 
$\Omega_b$.  
The cluster number density evolution depends significantly on $\sigma_8$ 
(see Figure \ref{fig2}). 
The two-dimensional likelihood in $(\sigma_8,\sum m_\nu)$ is shown in
Figure \ref{sig8}. Lower values of $\sigma_8$ favor a higher neutrino
mass.

 Our result of $\sum m_\nu$ $< 2.4$ eV, which is based on
the cluster number density evolution, is in good agreement with
bounds on neutrino masses from CMB (and other) measurements 
 \cite{elgaroy04,ichi} and
corroborates that evidence. Our limit indicates that effects from
the neutrinos on the evolution of galaxy clusters cannot be excluded and
also indicates that these effects should be taken into account in
the determination of cosmological parameters.
Based on our result and on the particle physics limit of 
$\sum m_\nu$ $>$ 0.04 eV, we find $\Omega_\nu$ is 
in the range of 0.1 \% to 5 \%.

\acknowledgments
We thank Arthur Kosowsky, Vladimir Lukash, and Sergei Shandarin
for fruitful discussions and suggestions.
We also acknowledge helpful comments from Neta Bahcall concerning the 
cluster mass function data. We acknowledge support
from CRDF-GRDF grant 3316, NSF CAREER grant AST-9875031,
DOE EPSCoR grant DE-FG02-00ER45824,
RFBR grants 01-02-16274 and 05-02-16658, and the Alexander von Humboldt
Foundation.

\end{document}